\documentclass[namedreferences]{kluwer}
\usepackage{graphics}      
     
\begin{document}       
\begin{article}       
       
\begin{opening}       

\title{HOW ARE EMERGING FLUX, FLARES AND CMES RELATED TO
       MAGNETIC POLARITY IMBALANCE IN MDI DATA?}  
       
\author{L.M. GREEN$^1$, P. D\'EMOULIN$^{2}$, C.H. MANDRINI$^3$,  
L. VAN DRIEL-GESZTELYI$^{1,2,4,5}$} 
        
\institute{$^{1}$  Mullard Space Science Laboratory, Univ. College London,
U.K. \email{lmg@mssl.ucl.ac.uk}     \\  
$^{2}$ Observatoire de Paris, section Meudon, LESIA
                 (CNRS), F-92195 Meudon Principal Cedex, France\\    
$^{3}$ Instituto de Astronom\'\i a y F\'\i sica del Espacio, IAFE,
                 CC. 67 Suc. 28, 1428 Buenos Aires, Argentina \\ 
$^{4}$ Centre for Plasma Astrophysics, K.U. Leuven,
                       Celestijnenlaan 200B, 3001 Heverlee, Belgium\\
             $^{5}$ Konkoly Observatory, Hungary}

\runningtitle{MDI MAGNETIC FIELD CHANGES}       
\runningauthor{L.M. GREEN ET AL.}       
       
\begin{abstract}               
In order to understand whether major flares or coronal mass ejections         
(CMEs) can be related to changes in the longitudinal photospheric magnetic 
field, we study 4 young active regions during seven days of their disc 
passage.  This time period precludes any biases which may be introduced in 
studies that look at the field evolution during the short-term flare or CME 
period only.  Data from the Michelson Doppler Imager (MDI) with a time cadence
of 96 minutes are used. Corrections are made to the data to account for area
foreshortening and angle between line of sight and field direction, and also 
the underestimation of the flux densities.  We make a systematic study of the 
evolution of the longitudinal magnetic field, and analyze flare and CME 
occurrence in the magnetic evolution.  We find that the majority of CMEs and 
flares occur during or after new flux emergence. The flux in all four active 
regions is observed to have deviations from polarity balance both on the 
long-term (solar rotation) and on the short term (few hours). The long-term 
imbalance is not  due to linkage outside the active region; it is primarily 
related to the east-west distance from central meridian, with the sign of 
polarity closer to the limb dominating. The sequence of short term imbalances 
are not closely linked to CMEs and flares and no permanent imbalance remains 
after them. We propose that both kinds of imbalance are due to the presence of
a horizontal field component (parallel to the photospheric surface) in the 
emerging flux.  
          
       
\end{abstract}       
\end{opening}       
       
\section{Introduction}       
            
Flares and coronal mass ejections (CMEs) are magnetic phenomena,        
thought to derive their energy from the coronal magnetic fields.         
However, reliable observations of the weak coronal fields are not yet        
available even though some attempts have recently been made        
\cite{lin00}.  Instead, indirect methods are used to determine        
information on the coronal fields by reconstruction using        
extrapolations, which have the photospheric flux distribution as the        
boundary condition, and use certain assumptions (e.g. that 
the field is force--free). The coronal fields are line tied at the        
photosphere, and it is then natural to investigate the photospheric        
field observations to search for signatures of flares and CMEs.         
Understanding the long-term evolution of the photospheric magnetic        
field leading up to the onset of CMEs and flares will help to constrain        
the theoretical models, and to identify the instabilities        
involved.  Still a confident signature relating the magnetic field        
evolution to event onset remains elusive.       
       
Vector magnetic field data enable us to study the magnetic shear in an 
active region (AR) by comparing the observed magnetic field direction 
with the one of the potential field computed with the same vertical 
field distribution.  Most studies (e.g.  \opencite{hagyard84} and 
references here after) have focused on the analysis of the magnetic 
shear across the photospheric inversion line of the vertical field 
component (because at this location the transverse field is best 
measured due to the absence of cross talk and Faraday rotation).  
Observations of M and X class flares have shown that the shear angle 
can decrease \cite{sakurai92}, increase \cite{wang94} or remain the 
same \cite{hagyard99,li00a} after a flare.  These different results, 
from flare to flare, are even present when data from the same 
instrument are analyzed by the same authors \cite{ambastha93,chen94}.  
Moreover, significant change in the magnetic shear is likely to be 
localized in the flaring part of the active region rather than on the 
inversion line \cite{li00b}.
              
In order to use vector data, the 180 degree ambiguity in the magnetic 
field direction must first be resolved.  This ambiguity is not a 
simple problem to deal with (a general solution is not presently 
known, see e.g.  \inlinecite{gary95} and references therein). Moreover,
magnetic fields crossing the inversion line in the opposite way to 
that of a potential field are known to exist, e.g.  in a prominence 
body \cite{bommier94} and also extending down in the prominence feet 
\cite{aulanier98}.  Furthermore, the errors in transverse vector field 
measurements are large, around 200~G, meaning that only data in high 
field strength regions can be considered as reliable.  However, even 
for large flares, flare ribbons are hardly observed within sunspots, 
indicating that energy release involves mostly the weaker fields 
around sunspots.  These large uncertainties on the transverse vector 
field are even reflected on global quantities, such as the free 
magnetic energy, and present observations do not have large enough 
spatial resolution and precise measurements to monitor the changes 
during a large flare \cite{klimchuk92}.  Then, the 180 degree 
ambiguity and the errors on the transverse field play an important 
role in the apparently mixed results of how the field changes around 
flare times.
       
Measurements of the longitudinal field component are more precise, 
because the errors are an order of magnitude smaller (20G).  
Studies of flares have shown that they occur frequently in magnetically
complex regions, such as $\delta$-spots (see e.g.~\opencite{gaizauskas98} 
and references therein) 
which are known to have highly sheared fields. Flares have also been 
linked to emerging 
flux  (see e.g.~\opencite{martres68}). Indeed, Martres et al. found that 
flares are often linked to two evolving polarities of opposite magnetic 
sign. They where able to follow the time evolution
only with the white light images. They found that flares are preferentially 
related  to polarities where one has a growing area while the other one 
is decreasing. They interpreted the white light area evolution as an increase 
in the magnetic polarity imbalance. However,~\inlinecite{harvey76} 
had difficulties finding a reliable polarity
evolution related to flares using magnetograms in both H$\alpha $ and 
Fe~I lines (in particular, they concluded ``flares do not seem to be 
associated with the areas where the largest flux changes occur'').
       
The most energetic flares, namely those with X-ray flux above 
10$^{-4}$ Wm$^{-2}$ (GOES X-class), are expected to be related to the 
largest evolution of the magnetic field configuration, and so they are 
the first to be investigated.  Most X-class X-ray flares occur in 
concert with CMEs (although this is not always the 
case,~\opencite{green02}) and these may offer a way to study the 
change in the AR fields related not only to the flare, but also to the 
CME.  Changes in the longitudinal magnetic field associated to X-class 
flares were investigated by~\inlinecite{wang02}.  They looked for 
changes in the magnetic field at the photospheric level around the 
time of the flare only, and found an impulsive change in each of the 
six cases studied which they proposed to be permanent.  They observed 
a lack of balance between the leading and following polarity; 
the  magnetic flux in the leading polarity increased from 10$^{20}$ Mx to 
10$^{21}$ Mx, while each event showed a decrease in the following 
polarity that could be between one order of magnitude lower and of 
similar magnitude as the corresponding increase in the leading 
polarity.   This effect was seen to be independent of the flare 
distance to the central meridian and was localized to a small area on the 
flaring neutral line in four out of the six cases, while it involved the 
full AR in the two others.  The authors suggest that the presence of a 
very inclined magnetic field could explain these deviations from polarity balance 
(due to projection effects).  Still they needed three 
different mechanisms to, at least qualitatively, explain their 
observations (see their Table~2).
        
When using longitudinal data to study CME onsets, it may be more useful 
to look to the long-term evolution of the magnetic field in the period 
leading up to the CME to find clues on the instability which results 
in the eruption, since a long-term shearing of the magnetic 
configuration is expected in most theoretical models.  It is also 
important to investigate the magnetic field variations outside the 
flare and CME period to infer if the changes associated to the flare 
and CME are relevant or not.  

On the short term, since the coronal instability initiating the CME 
is not able to alter significantly the vertical 
magnetic field distribution at the photosphere, we do not expect to observe 
important magnetic changes at the photospheric level as a result of a CME.  
However, we may see an evolution of the longitudinal field indicating, 
for example, the role of new flux emergence.  \inlinecite{lara00} used 
MDI data to search for a photospheric signature in a time period of a 
few days, before, during and after 8 CMEs.  They found that the 
observed magnetic flux 
in the entire active region showed no obvious change associated to the 
CME on the short term, but that flare associated CMEs occurred during 
the maximum phase of magnetic flux emergence.  In fact, many works have related 
flare and CME activity to emerging flux; for example, for flares most 
of the works cited in the second and fourth paragraphs of this section, while for 
CMEs we refer to \inlinecite{feynman95}, \inlinecite{plunkett97}, 
\inlinecite{tang99}, and \inlinecite{wang99}.  The usual requirement being 
that the emerging magnetic flux is oriented favourably to allow reconnection 
with the pre-existing field.
              
In this paper we present observations of the evolution of the 
longitudinal magnetic field of 4 emerging active regions as they cross 
the solar disc.  Magnetic data from the MDI instrument onboard SOHO 
are used to determine the total observed flux of both magnetic 
polarities in each AR.  
We study whether any deviation from polarity balance
is an instrumental or geometrical 
artifact.  Section~\ref{sect_mdi} describes the Michelson Doppler 
Imager and Section~\ref{sect_anal} details our analysis.  
Section~\ref{sect_results} presents the ARs chosen and their flare and CME
activity. 
Section~\ref{sect_flux_evol} details the time evolution of the deviation from 
polarity balance.  In Section~\ref{sect_im_orig} we 
discuss the possible origins for these deviations.  
We conclude in Section~\ref{sect_con} about the absence
of a confident signature  which relates CME or flare 
occurrence to deviation from polarity balance. We rather relate these 
imbalances to flux emergence.

\section{Overview of MDI}             
\label{sect_mdi}       
       
The Michelson Doppler Imager (MDI,~\opencite{scherrer95}) takes five        
narrowband (94 m\AA) filtergrams at different positions along the Ni I        
6767.8 \AA~absorption line, formed in the mid-photosphere.  MDI then        
computes Doppler velocity and continuum intensity from the        
filtergrams.  The velocity calculation refers to an onboard lookup        
table constructed from synthetic line profiles and measured filter        
transmission profiles.  Waveplates allow the right (RCP) and        
left-handed circular (LCP) polarization signals to be measured, and        
the longitudinal magnetic flux density is given by the difference in        
Doppler shifts of the RCP and LCP values.  The spatial resolution of        
the full disc magnetograms is $\approx$4 arcsec (pixel size is        
$\approx$2 arcsec).  MDI returns the flux density averaged over the        
pixel field of view and from this the flux within a required area can        
be found.       
       
As with any magnetograph, MDI has some instrumental effects.  Data with 
low continuum intensities are expected to be related to strong field 
regions and, therefore, they should have high flux densities.  Lower 
than expected flux densities may be produced by mixed polarities 
within the resolution element, as MDI returns the flux density 
averaged over the pixel field of view.  However, it has also been 
observed that sunspot umbras with very low continuum intensity values 
sometimes return low flux densities instead of the high values 
expected.  Further investigation showed that this is produced by a 
failure in the onboard algorithm when the lookup table saturates 
\cite{liu01}.  Indeed, very low continuum intensities can result in 
the Ni I line profile almost disappearing.  This produces corrupted 
pixels and the observations show the unexpected result of the flux 
densities in the centre of the sunspot umbra being much lower than 
that in the surrounding, outer pixels.  These corrupted pixels can be 
identified by plotting the pixel continuum intensity against flux 
density.  A tail of data points, where counter-intuitively the lowest 
intensity pixels have low magnetic flux densities, is formed when 
corrupted data are present.  \inlinecite{liu01} also found, using the 
Harvard-Smithsonian Reference Atmosphere model \cite{giggerich71} to 
compute line profiles, that MDI underestimates flux densities by as 
much as 30\%, with the largest underestimation applying to the higher 
flux densities.    
              
\inlinecite{berger02} analyzed co-temporal and co-spatial MDI data 
with data from the Advanced Stokes Polarimeter (ASP) to compare the 
flux densities measured by each instrument.  They found that MDI 
systematically measured lower flux densities than did ASP.  MDI 
underestimated the flux densities in a linear way for MDI pixel values 
below $\approx$ 1200G by approximately  a factor 1.45.  For flux 
densities higher than 1200G the underestimation became strongly 
non-linear and the magnetic field measured by MDI saturated at 
$\approx 1300$~G, whereas for ASP the corresponding field increased. The 
underestimation appears to be inherent to the MDI calibration and does 
not result from the different instrument resolutions.
       
\section{Data Analysis}                           
\label{sect_anal}       
       
This study uses the full disc level 1.5 MDI magnetograms.  These data 
are the average of 5 magnetograms with a cadence of 30 seconds and a 
noise error of 20~G per pixel~\cite{scherrer95}.  They are constructed 
once every 96 minutes.  The error in the flux densities per pixel in 
the averaged magnetograms is then $20/\sqrt{5} = 9$~G, and 
each pixel has a mean area of $1.96$~Mm$^2$ (the exact area depends on 
the Sun-Earth distance).  

In order to find the AR flux, a polygonal contour 
defined by eye is fitted around the AR, taking as the boundary the 
sharp flux density change between the young AR and the network field. 
Within this region the flux is summed.  In this way we minimize the contribution 
of the background field not related to the AR. The number of sides to the
polygon is determined by the complexity of the AR shape, but is typically
close to 20. The flux evolution of 
the ARs cannot be followed to the solar limb as the data become too 
distorted within approximately 25 degrees of the limb.
       
The $\pm9$G noise in each MDI pixel introduces an error into the AR 
flux measurements, since each AR covers around 5000 pixels.  For an AR 
with a total flux of the order of $10^{22}$Mx, the noise introduces an 
error of around 0.1\%.  In fact, the main error in the AR flux 
measurement arises from the definition of the polygonal region within 
which the flux is found.  There will be included in the region small 
flux elements of the background field.  The change in the total flux, 
from one flux measurement to the next, associated to these flux 
elements moving across the polygon boundary produces a modification of 
the imbalance of the order of
1\% of the AR flux as the flux in these elements is low (of the order 
$10^{20}$Mx). Whenever these elements enter or cancel 
with the opposite polarity inside the defined polygon, they produce 
short term (on time scale of hours) deviations from polarity balance. 
The tiny fluctuations present in the curves in  Figure~\ref{fig_flux_evol} 
include such effects (curves are not smoothed, so they have a time 
resolution of 96~min). On the long-term, 
the presence of a dominant background polarity surrounding the AR has 
a cumulative effect (with or without cancellation with the AR field) 
since more and more background field is included in the ever increasing 
AR area. The magnitude of this
effect is more difficult to quantify, but we note that the main objective
is rather to study the possible link between CMEs, flares and  the short-term 
imbalance. It is possible to derive an upper bound estimate of this systematic effect
at CMP, Figure~\ref{fig_flux_evol} shows that  it is below $10^{21}$~Mx for the
four ARs studied (and that other contributions to the imbalance are present: see 
Section~\ref{sect_im_orig.6}).

In order to study the evolution of the magnetic field, the MDI data        
must be corrected for the instrument calibration effects and        
geometrical distortions.  These include: 
 \begin{itemize}
 \item  Correcting for the angle between the magnetic field direction 
and the observer's line of sight.  In this case we assume that the 
field is radial at the photosphere and correct for the reduction by 
$cos\varphi$ that the flux density experiences as the AR moves 
toward the limb, where $\varphi $ is the angular distance of the pixel from the
central meridian of the image.
 \item The foreshortening of the AR area with distance from central        
meridian passage (CMP).
 \item Underestimation of the flux density.              
 \item Identification of corrupted pixels.         
 \end{itemize}       

Following the results of~\inlinecite{berger02}, we first correct all 
the magnetic field measurements by a factor $1.45$ (because they have 
found $B_{\rm MDI} \approx 0.69B_{\rm ASP}$).  When MDI is measuring 
flux densities above approximately 1200~G, the relationship between the 
data from the two instruments becomes non-linear and the flux recorded 
by MDI appears to saturate.  To evaluate this effect we make a second 
flux measurement in which the field above 1200~G is multiplied by a 
factor $1.9$ (rather than $1.45$, as before).  This factor 1.9 is 
introduced as an upper bound to the MDI underestimation, this comes 
from the largest difference between ASP and MDI measurements found 
by~\inlinecite{berger02} when the MDI data saturates (see their Fig.  
2).  The change in flux, $\Phi$, after the correction for both linear 
and non-linear response of MDI is then given by:
  \begin{equation}       
   \label{flux-corr}       
   \Phi_{corr.} =  1.45 (\Phi + 0.3 \Phi_{B>1200~G})        
  \end{equation}       
So the further non-linear correction of MDI response is relatively 
small after taking into account the linear correction (e.g.  even if 
half the flux is above $1200$~G, the non-linear correction is upward 
bounded by a 15\% increase).
       
\begin{table}       
\begin{tabular*}{\maxfloatwidth}{lllll} \hline        
NOAA number   & 7978 & 8086 & 8100 & 8179   \\ \hline          
Central meridian & 7 Jul. 1996 & 18 Sep. 1997 & 2 Nov. 1997 & 15 Mar.        
1998 \\       
Leading polarity      & positive & positive & negative & negative \\       
Latitude (degrees)    & S10      & N27      & S19      & S22      \\        
\hline       
\end{tabular*}       
\caption{Active region information.}       
\label{ar_table}       
\end{table}

In order to search for any association between eruptive events and 
changes in the longitudinal field, CMEs from each AR were identified 
using white light data from the C2 (2-6 solar radii) and C3 (4-30 
solar radii) Large Angle and Spectroscopic Coronagraphs 
(LASCO,~\opencite{brueckner95}) onboard SOHO.  The CME times given in 
this paper are the times when each CME was first observed in C2 and 
not the onset time, which is harder to determine.  Lower coronal 
signatures~\cite{thompson98} were used to locate the source region of 
each CME using data from the Extreme-ultraviolet Imaging Telescope 
onboard SOHO (EIT,~\opencite{delaboudiniere95}).  Major flares, that 
is those with GOES class M (10$^{-5}$ Wm$^{-2}$) or X (10$^{-4}$ 
Wm$^{-2}$), were identified for each region during their disc passage 
and correlated with Solar Geophysical data reports on H$\alpha$ solar 
flares.  Flare times shown in Figure~\ref{fig_flux_evol} are the start 
times of the X-ray flares.  Four emerging active regions were studied 
from 1996 to 1998, they had the NOAA numbers 7978, 8086, 8100 and 
8179.  Information on these ARs are detailed in Table~\ref{ar_table}.

\section{Active Regions Studied}       
\label{sect_results}       
       
Each AR was studied during its first disc passage.  After proceeding 
as summarized in Section~\ref{sect_anal}, it was found that corrupted 
pixels were not a significant problem for any of the ARs.  ARs 7978, 
8086 and 8100 have no corrupted pixels during the time periods 
studied.  In AR 8179 there is a maximum of 3 corrupted pixels in any 
one magnetogram, first seen in the positive polarity on 16 March 1998 
14:24 UT and until 18 March 1998 16:03 UT.  The corrupted pixels 
account for a loss of approximately 1\% of the total flux in the 
positive polarity.
       
{\bf AR 7978} was born onto the solar disc with positive leading 
polarity in July 1996, into a region of dominantly negative background 
field in the southern hemisphere.  Flux emergence was first observed 
on 4 July 1996 (see~\opencite{demoulin02}, Section 2.2).  The 
long-term evolution and activity of this AR has been well studied by 
several authors including \inlinecite{dryer98}, 
\inlinecite{vandriel99}, \inlinecite{mandrini00} and 
\inlinecite{demoulin02}.  The new flux emergence occurred close to the 
time of CMP and so the evolution is monitored only in the western 
hemisphere.  During this disc transit the AR produced 4 slow CMEs and 
3 major flares (GOES class M1.4, X2.6 and M1.0).  The X-class flare 
and second M-class flare are associated to the second and third CMEs 
respectively, and so these flare times reflect also the CME initiation 
time.
       
\begin{figure}       
 \resizebox{\hsize}{!}{\includegraphics{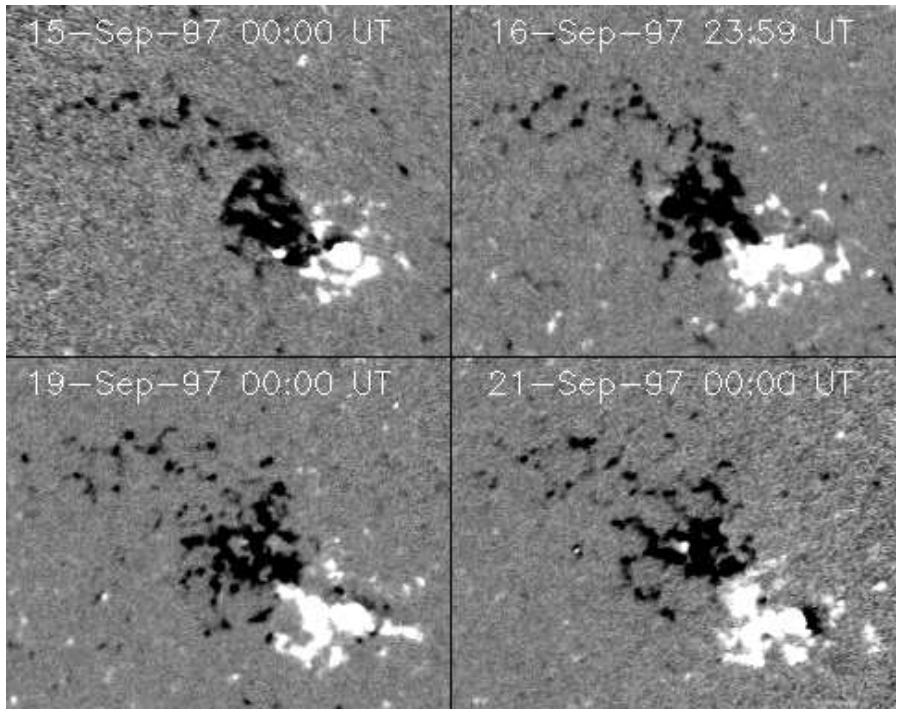}}       
\caption{MDI data showing the longitudinal magnetic field        
evolution in AR 8086.  White represents positive field (toward        
observer) and black represents negative field (away from observer).        
The images have been corrected for area foreshortening which occurs        
away from central meridian passage on 18 September 1997. The size of        
the boxes is 348$\times$276Mm.}       
 \label{8086_mdi}       
\end{figure}       
       
{\bf AR 8086} was born in the northern hemisphere with positive 
leading polarity. The region
formed sunspots when it was close to the east limb.  The new flux 
emerged into the
magnetic fields of a decaying AR.  Sunspot decay and flux dispersal 
were observed during the disk passage
(Figure~\ref{8086_mdi}).  The AR showed flare activity but produced no M 
or X-class flares.  Transient emission in the low corona and changes in
the magnetic topology (as detected with
EIT/SOHO) showed that 2 CMEs 
occurred during the disc passage, observed in LASCO/C2 on 13 September 
1997 11:33 UT and 22 September 1997 07:29 UT.  The AR was very close 
to the limb on both occasions and so these events do not occur within 
the time period in which the AR flux is being measured.  AR 8086 
allows us to study the flux evolution in a region during a period of no flare 
or CME activity.

\begin{figure}       
 \resizebox{\hsize}{!}{\includegraphics{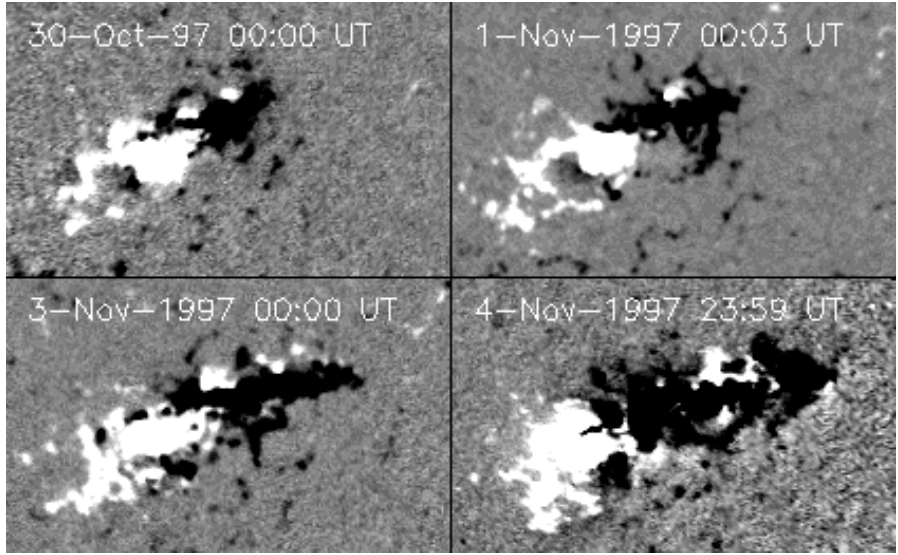}}       
\caption{Same as Figure 1, but for the flux evolution in AR 8100.  The        
central meridian passage occurs on 2 November 1997.  The size       
of the boxes is 305$\times$189Mm.}       
 \label{8100_mdi}       
\end{figure}       
       
{\bf AR 8100} is a southern hemisphere AR with negative leading        
polarity that first appeared on 28 October 1997.  A revival of        
strong flux emergence was observed beginning on 2 November 1997        
(Figure~\ref{8100_mdi}).  AR 8100 was highly flare and CME        
productive \cite{delannee00,green02b} producing 5 M-class flares, 2        
X-class flares and 15 CMEs between 2 and 9 November        
1997.  The evolution of the magnetic field in this AR is detailed        
in~\inlinecite{green02b}.

\begin{figure}       
 \resizebox{\hsize}{!}{\includegraphics{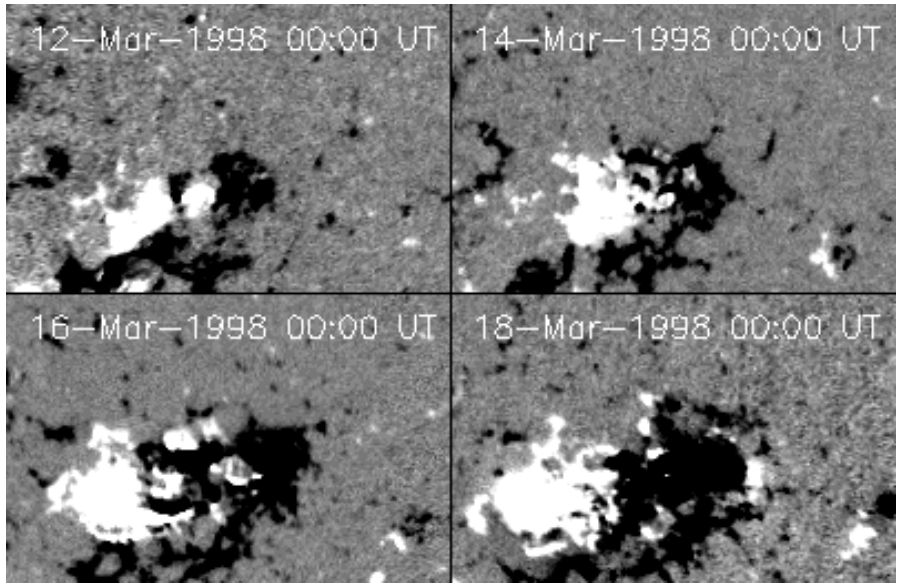}}       
\caption{Same as figure 1 but for the flux evolution in AR 8179.  The        
central meridian passage occurs on 15 March 1998.  The size of the        
boxes is 290$\times$189Mm.}       
 \label{8179_mdi}       
\end{figure}        
       
{\bf AR 8179} emerged into the southern hemisphere and had a negative        
leading polarity.  Figure~\ref{8179_mdi} shows the formation of the AR        
in MDI data.  New flux emerged late on 13 March 1998 into a        
pre-existing bipole in a complex of decaying active regions.  The        
negative leading polarity was observed to be more dispersed than the        
following positive, opposite to the general rule.  Two CMEs were        
produced which were first observed in LASCO/C2 on 15 March 1998 19:31        
UT and 18 March 1998 11:35 UT, 4 M-class X-ray flares occurred during        
the same period, the second of which is associated to the first CME.

\section{Magnetic Flux Evolution}        
\label{sect_flux_evol}       
              
Figure~\ref{fig_flux_evol} shows the flux evolution of the 4 active 
regions after corrections for area foreshortening, reduction in flux 
density due to the angle between observer and (assumed radial) 
magnetic field direction, and MDI underestimation of the field.  The 
time that each region crosses the central meridian is indicated by a 
thick dashed line.  CME times, as first observed in LASCO/C2, and 
start times of major flares are marked by solid and dash-dotted lines, 
respectively.  The upper plot for each AR shows the evolution of the 
positive (continuous line) and negative (dashed line) fluxes after the 
correction by factor 1.45 for flux underestimation.  This is applied 
to all flux density values.  The lower plot shows the difference 
between observed flux in the leading polarity and following polarity
as follows. The continuous line gives the polarity imbalance after the 
factor 1.45 correction for all flux densities.  The 
dotted line, however, shows the polarity imbalance after correction by 
factor 1.45 for flux densities below 1200 G, and also the additional 
correction by factor 1.9 for flux densities above 1200G. 
The lower plots of Figure~\ref{fig_flux_evol} 
make  more visible the deviations from polarity 
balance and also the contribution of the two corrections.
       
The new flux emergence into each region is clearly observed and in all        
regions the deviation from polarity balance has
variations on both the long and        
short-term, as we describe in more detail below.  The long-term        
imbalance increases with distance from CMP and        
the polarity which is closest to the limb always appears to have more        
observed flux than the polarity closer to disc center.       
       
For {\bf AR 7978}, the  observed magnetic flux of both polarities grows 
monotonically with time (Figure~\ref{fig_flux_evol}).  The
deviation from polarity balance 
is first observed approximately one day after central meridian passage 
when more positive (leading polarity) than negative (following 
polarity) flux is measured.  Later, the deviation from polarity balance
has a long-term 
monotonical increase with time, as the AR rotates toward the western 
limb, with only a weak modulation superimposed on top.  The first CME 
(seen in LASCO/C2 on 8 July 1996) occurs close to the time when the 
deviation from polarity balance
started to grow ($\approx 10^{21}$~Mx in 6~hours, at the time 
of new flux emergence).  However, the imbalance stops 
increasing, and even slightly decreases after the X2.6 flare on 9 July 
1996 associated to the second CME (decrease of $\approx 
10^{21}$~Mx).  This lasts approximately 5 hours before the 
imbalance starts again on its long-term increase.  The two M-class 
flares (and so the third CME which is associated to the second M 
flare) appear not to be accompanied by any changes on top of the 
long-term increase. Conversely, short-term deviations from polarity balance 
are observed without the occurrence of a major flare (although one may 
be coincident with a C-class flare), e.g., notice the increase of 
$\approx 10^{21}$~Mx at the end of 10 July and the decrease of 
$\approx 2\times10^{21}$~Mx by midday on 11 July, both lasting a few hours.     
       
\renewcommand{\topfraction}{0.9}        
\begin{figure}       
 \resizebox{\hsize}{!}{\includegraphics{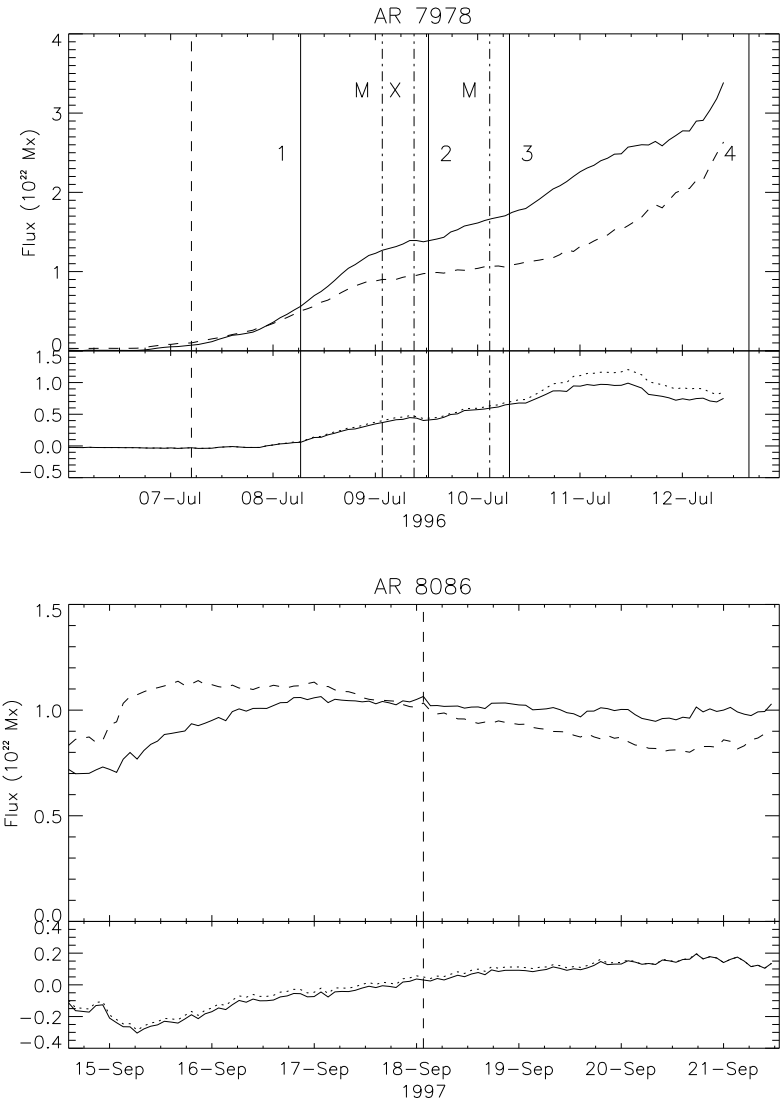}}       
 \caption{Flux evolution for active regions 7978 and 8086. The upper        
plot for each indicates the positive (continuous curve) and negative        
(dashed curve) flux evolution.  The vertical thick dashed line        
represents the time of central meridian passage of each region.         
CME times as first observed in LASCO/C2 are shown by        
solid vertical lines and times of major solar flares are shown by        
dash-dotted lines.  The flux has been corrected for the geometrical        
effects of area foreshortening and angle between magnetic field        
direction and observer, and linear underestimation (factor 1.45) by        
MDI.  The lower plot for each active region shows the flux difference        
(or polarity imbalance) between leading and following 
polarity after correction for linear        
underestimation (solid line), and also for linear and        
non-linear underestimation (dotted line).}       
\label{fig_flux_evol}       
\end{figure}        
       
\addtocounter{figure}{-1}       
\begin{figure}       
 \resizebox{\hsize}{!}{\includegraphics{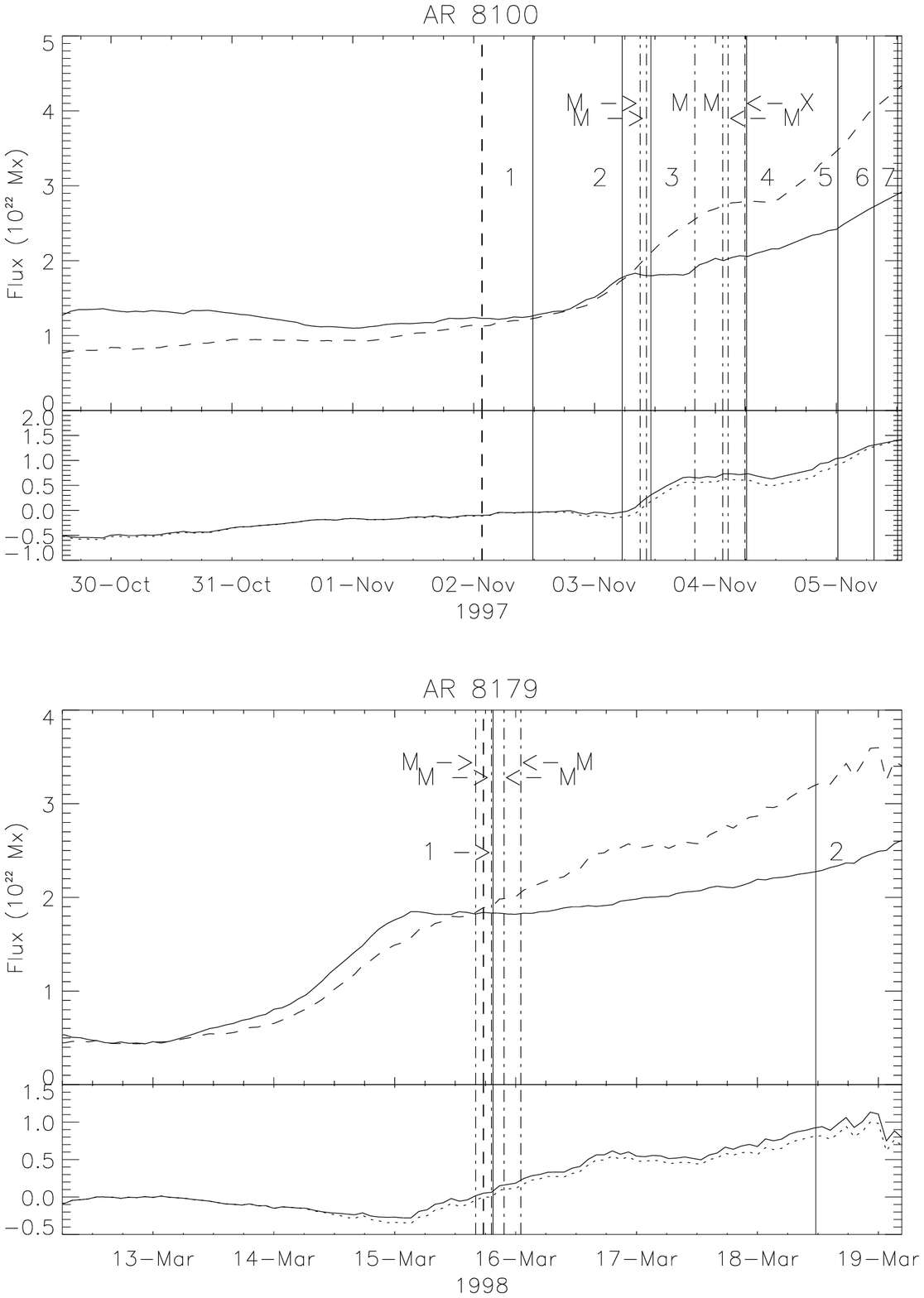}}       
\caption{ {\it continued}. Flux evolution in active regions 8100 and        
8179.}       
\end{figure}       
       
In {\bf AR 8086} there is a clear imbalance in favour of the negative 
(following) polarity in the eastern hemisphere and in favour of the 
positive (leading) polarity in the western hemisphere 
(Figure~\ref{fig_flux_evol}).  The polarities have equal magnitudes on 17 
September around 18:00 UT, just prior to central meridian passage.  
The CMEs from this AR occur very close to the limb, and so it is not 
possible to look for deviations from polarity balance 
associated to the eruptive events.  Except in the early times, the 
mean flux of the polarities of this AR is steadily decreasing with 
time.  The imbalance shows a long-term 
trend, with the limbward 
polarity measuring more flux than the other, with smaller variations 
on top ($\leq 0.5\times10^{21}$~Mx).
       
Fluxes in {\bf AR 8100} are imbalanced in a way that the positive 
(following) polarity dominates when the AR is in the eastern 
hemisphere, but the imbalance reduces as the AR approaches CMP 
(Figure~\ref{fig_flux_evol}).  Just after CMP the polarities have the same 
flux and remain this way during 1 day when, coincident with new flux 
emergence, the negative (leading) polarity begins to dominate.  This 
imbalance increases as the AR flux increases and as the AR moves toward the 
western limb.  AR 8100 is the source of 15 CMEs and 7 major flares 
between 2 and 9 November 1997 (see Table 1 of \opencite{green02b}).  
For 2 of the CMEs it was not possible to determine with confidence the 
time of first observation in LASCO/C2 (so they are not represented in 
Figure~\ref{fig_flux_evol}); they are observed after CME 1 and 2 
respectively.  The first CME (on 2 November) occurs around the start 
of the revival of flux emergence.  CME number 2, which occurs on 3 
November, is observed to occur before the time when the 
deviation from polarity balance
in the AR starts to grow, while CME 3 is just in the growing phase of 
the imbalance.  The imbalance grows by $\approx 7\times10^{21}$~Mx in 
12~hours with only one CME associated flare (CME 3) and two more 
flares at the beginning and end phase!  The first M-class flare on 4 
November appears to be accompanied by a decrease in the 
imbalance 
(the positive flux increases by $\approx 2\times10^{21}$~Mx in 
6~hours).  The X-flare (4 November 1997 05:52 UT), which is associated 
to CME 4, is followed by a decrease of the imbalance
($\approx 10^{21}$~Mx in 6~hours).  For CMEs 5 and 6 there are no related 
changes in the imbalance and it continues to grow monotonically.  
CME 7 is too close to the end of the plot to look for changes.
      
Flux measured in {\bf AR 8179} is balanced until the time of the new 
flux emergence in the eastern hemisphere at which time the positive 
(following) polarity becomes dominant (Figure~\ref{fig_flux_evol}).  The 
flux becomes balanced just before central meridian passage, and then 
in the western hemisphere, the negative (leading) polarity becomes 
dominant.  The first CME and associated M-class flare are seen to 
occur with an increase in the deviation from polarity balance
(this happens close to 
the CMP and lasts less than 3 hours).  The second CME (on 18 March) 
shows no significant change of flux evolution on top of the long-term 
trend.  Still, earlier, significant changes are present without any 
flare or CME association (e.g. the two imbalanced flux increases by 
$\approx 2\times10^{21}$~Mx in 6~hours on 16 and 17 March).

\section{Origin of the deviation from polarity balance}       
 \label{sect_im_orig}       
       
\subsection{MDI non-linear response}       
 \label{sect_im_orig.1} 
An a priori possible origin of the above described deviation from 
polarity balance is 
the non-linear response of MDI. For example, if magnetic flux evolves 
from sunspot-like to plage-like, a false field increase can be 
detected by MDI.  In the study of~\inlinecite{berger02}, this became 
important for flux densities above 1200G as MDI underestimates the 
field in a non-linear way (Section~\ref{sect_mdi}).  For the ARs 
chosen in this work, we suppose that the linear response breaks 
down as in the AR studied by~\inlinecite{berger02}
 (see Section~\ref{sect_anal}).
In ARs 7978, 8100 and 8179 flux 
densities above 1200G are observed in each polarity.  In AR 8086 flux 
densities above 1200G are only seen in the positive polarity.  The 
difference in flux between the leading and following polarity, after 
correction for flux densities larger than 1200G are shown in the 
lower plots of Figure~\ref{fig_flux_evol} by the dotted line.  This 
correction has the effect of increasing the long-term imbalance in AR 
7978, but decreases the long-term imbalance in ARs 8100 and 8179, 
while it has a negligible effect in AR 8086.  As expected from 
Eq.~(\ref{flux-corr}), Figure~\ref{fig_flux_evol} shows that correcting 
for the non-linear underestimation has a minor effect (even when we have 
done an upper bound correction).  We conclude that the non-linear 
response of MDI is not the source of both the long-term and 
short-term lack of balance between opposite polarities.

\subsection{Flare and/or CME origin}       
 \label{sect_im_orig.2} 
        
Studying the flux evolution during the entire disc passage of an AR, 
allows any relation between changes in flux and times of flares and 
CMEs to be investigated systematically. On the short-term 
($\approx 6$~hours), some flares and CMEs show a deviation 
from polarity balance 
around the flare/CME time, for example the 9 July 1996 09:05 UT X2.6 
flare.  These changes are typically of the order of 
$1-2\times10^{21}$~Mx.  In our data set, 4 flares (and their 4 
associated CMEs) show such related short-term imbalance but the 
same numbers of events show no significant changes (while others, 4 
flares and 2 CMEs, were occurring too close to CMP to expect any 
significant change).  We conclude that the association between 
the polarity imbalance and the 
occurrence of major flares and CMEs is not systematic.  In the cases 
where we do have a deviation from polarity balance, this effect has a 
typical duration $\approx 6$~hours.
      
Complementary work carried out by \inlinecite{wang02} led to the 
proposal that deviations from polarity balance associated to flares 
are permanent, although at most their observations cover the time 
period of 4 hours after flaring.  They look at 6 ARs with a cadence of 
one minute (compared to our 96 minute cadence).  In the fully 
comparable cases, short-term effects were seen corresponding to a 
change lasting 1 hour in AR 9591, and only 30 minutes in AR 9672.  The 
flare related changes are very small, of the order of $10^{20}$ - 
$10^{21}$ Mx.  Changes of $10^{20}$ Mx are hard to distinguish from 
noise in this study, where the total AR flux is at least one order of 
magnitude larger than that measured by \inlinecite{wang02}.  The total 
flux measured by these authors is much smaller than in our study 
because, except in one case (AR 9672), they defined a sub-region which 
covers only a small part of the studied AR.  Having such a defined 
sub-region located exactly on the same portion of the AR during the 
time period studied, is certainly an important challenge!  The results 
are also very dependent on the location of the sub-region in the 
studied AR, as found by \inlinecite{lara00}.  We fully confirm the 
difficulties of such an attempt and, finally, we rely only on the flux 
evolution of the full AR (even in this case the AR is never fully 
isolated because it is surrounded by network field, and this is indeed 
the main source of noise in the deduced magnetic flux, see 
Section~\ref{sect_anal}).

The cadence in our case is 96 minutes, restricting our determination 
of short-term changes, but allowing us to study the permanent and 
long-term polarity imbalance more precisely.  
However, no permanent 
imbalance is observed to be related to flares as 
proposed by \inlinecite{wang02}.  We do see imbalance associated to 
some flares, but our systematic study shows that imbalances of 
similar magnitude can also occur outside flare and CME times.  We 
conclude that flares and CMEs are not closely related to the 
observed changes in magnetic polarity balance.

\subsection{Global connectivities}       
 \label{sect_im_orig.3}               
The results presented in Figure~\ref{fig_flux_evol}  show that 
the polarity imbalance increases with the distance from the CMP. 
For the four ARs, the imbalance is minimum near the CMP.
It shows that the main source of the imbalance is 
not due to global connectivities which link part of the AR flux 
to outside distant flux (as proposed by \opencite{choudhary02}). 
       
\subsection{Flux emergence}       
 \label{sect_im_orig.4} 
The nature of the imbalance corresponds to what is expected 
from the presence of an east-west horizontal field component which 
links the two AR polarities as follows. Such a horizontal component 
gives a stronger 
contribution to the line-of-sight flux as the AR moves away from the 
central meridian (where it nearly vanishes), just because of a 
geometrical effect. This increase of imbalance  
with the distance from the CMP is clearly present in Figure~\ref{fig_flux_evol}.
Moreover, the long-term evolution of the polarity imbalance is associated with 
an increase of flux in both polarities, so with new emerging 
flux (Figure~\ref{fig_flux_evol}). 
The east-west horizontal component of this new flux introduces a 
longitudinal component in the observed field component (there is also 
a much smaller contribution from the north-south component, see the 
end of Section~\ref{sect_im_orig.6}). With a static magnetic configuration, 
the polarity imbalance created by the east-west component would 
have a sinus dependence 
on the distance from the CMP.  In an evolving magnetic configuration, 
only the reversal
at CMP, and an average increase with the distance from CMP remain as 
shown in Figure~\ref{fig_flux_evol} and described in 
Section~\ref{sect_flux_evol}. 
    
It has been suggested previously that CMEs occur during the maximum 
phase of new flux emergence (e.g.  \opencite{lara00}).  We observe 
CMEs during all stages of flux emergence, i.e.  from first emergence 
to days after but we do also observe important flux emergence without 
CMEs.  In fact, it is likely that CME occurrence depends on the 
stability of the overlaying coronal field and its interaction with the 
emerging flux.  Indeed, from a wide range of models, new emerging 
flux (and/or cancellation of flux and/or increasing magnetic stress) 
is not expected to systematically produce a flare or a CME (e.g.  
\opencite{isenberg93}, \opencite{amari96}, \opencite{antiochos99}, 
\opencite{lin01}).  So finally, it is physically grounded that we 
find the deviations from polarity balance related to flux emergence,
without any systematic link to flare or CME.
  
\subsection{Implication of the radial field hypothesis}       
 \label{sect_im_orig.5} 
It is also worth remembering that our above results are obtained after 
correcting the magnetic flux using a radial field hypothesis (see 
Section~\ref{sect_anal}).  If we do not make the corrections 
implied by this assumption, 
the observed flux of the polarity which is closer to the limb is 
systematically lower (the opposite of what is shown in 
Figure~\ref{fig_flux_evol}~!).  Indeed, for an AR with a nearly constant 
magnetic flux (e.g.  AR 8086) and without the radial correction, we 
find that the longitudinal magnetic flux of both polarities has an 
approximate cosine dependence with the  distance from CMP; the leader 
polarity flux peaks earlier than the following polarity flux (with a 
time difference proportional to the east-west AR  size).  
Multiplying each pixel of the magnetogram by a factor $cos\varphi 
^{-1}$ (Section~\ref{sect_anal}) corrects most of this projection 
effect (because the magnetic field is dominantly vertical at the 
photosphere in the spots or in the network).

Still, we attribute above, most of the remaining 
deviation from polarity balance to 
the east-west horizontal component!  The correction by a factor 
$cos\varphi ^{-1}$ certainly does not apply to such component, but we 
are unable to separate the contributions of the vertical and 
horizontal components to the longitudinal field.  Therefore, we choose 
to do the projection correction on the component which has the 
strongest flux.  This implies that the polarity imbalance coming from the 
horizontal component are over-estimated (with a factor increasing with 
the distance from CMP).  For an AR at a latitude around $20$ degrees, and 
3 days before or after CMP, this over estimation is lower than a 
factor 1.5.  It is also worth noticing that this correction has only a 
long-term effect (linked to the solar rotation), and so it cannot be 
linked to the short term ($\approx 6$~hours) 
deviations from polarity balance.
        
\subsection{Other possible contributions}       
 \label{sect_im_orig.6} 

   Aside from an horizontal contribution due to emerging flux, horizontal 
fields are also known to exist in the penumbra of sunspots.  If sunspots 
are axisymmetric this would not produce a deviation from polarity balance 
(because of the averaging in the total flux).  However, sunspots are 
usually not symmetric.  Moreover, \inlinecite{howard91} deduced from 
the east-west flux difference that leading and following photospheric 
magnetic fields are inclined toward each other by about 16~degrees.  
Howard found this result for both growing and decaying regions.  Then 
we expect a deviation from polarity balance in decaying ARs with the same 
longitudinal dependence (but with a weaker magnitude) as for growing ARs.
AR~8086 is one such example. 

Exact measurements of the flux in each polarity, with only an 
east-west contribution from the horizontal field, would result in 
fluxes having equal magnitude at the time when the AR crosses the 
central meridian.  Figure~\ref{fig_flux_evol} indicates that at CMP, 
most notably for AR 8086 and 8100, the fluxes of opposite polarity
are not equal.  This may 
result from the inclusion of some of the background field or 
pre-existing bipolar fields in the flux summations; in particular, 
even if we isolate the AR the best we could, network field present in 
the AR area cannot be removed from the flux computation.
          
Another possible contribution comes from a north-south oriented 
horizontal field component in the AR.  Both ARs 8086 and 8100 show a 
much higher inclination toward the equator than ARs 7978 and 8179, 
which are more aligned in the east-west direction.  However, the 
north-south field component contributes with the correct sign only in 
AR 8100 (to explain qualitatively a slight positive dominance at CMP), 
while for AR 8086 only a dominant positive background flux or large 
scale connections are able to explain the slight positive dominance at 
CMP.  Nevertheless, the dominant effect for the polarity imbalance is the 
distance in longitude of the AR from CMP, so the contribution of the 
east-west horizontal field.

\section{Conclusion}       
\label{sect_con}       
       
This study seeks to answer whether short-term, or permanent changes in 
the longitudinal photospheric magnetic field can be associated to 
major flares and CMEs by studying ARs during their disc passage.  
Since MDI, like any magnetograph, has limitations, certain corrections 
must be made to the data before they can be used.  As discussed in 
Section~\ref{sect_mdi}, in pixels recording a very low continuum 
intensity we can have corrupted values of the flux density.  In pixels 
which are not corrupted, MDI will underestimate the flux density 
either in a linear or non-linear way~\cite{berger02}.  MDI data in 
this study have been corrected for these effects.  In the ARs studied, 
corrupted pixels were not a significant problem.  They were observed 
in one of the emerging ARs and produced at most a loss of 1\% of the 
flux in the corresponding sunspot.  Corrections have been applied to 
the data to take into account the geometrical effects of area 
foreshortening and the inclination of the field away from the line of 
sight, and also for the underestimation of the field when the response 
of MDI is linear and non-linear.  However, we find that this 
non-linear underestimation is not the source of the 
deviation from polarity balance in the ARs.
       
Even after these corrections, the magnetic flux in both polarities is        
observed to have a long-term imbalance (with a disk-passage time scale). 
This lack of polarity balance increases with distance from disk centre, 
the stronger 
flux being associated to the polarity which is farther from disk centre.
At central meridian passage the fluxes become approximately equal.  
The nature of the imbalance is that expected from the presence of a 
horizontal field component linking both polarities of the AR.  The ARs 
studied were young, in the early stages of formation, so that the 
presence of such horizontal field component is expected (in particular 
because the emergence of new flux implies that the apex of the flux tube 
crosses the photosphere).

Furthermore, superposed on the longitude dependent long-term 
polarity imbalance, we have found several enhancements or decreases of the 
imbalance on the short-term (with a typical time scale of 6~hours).  
Some occur around the time of a major flare and/or a CME associated to the 
same AR, but without a systematic timing between the 
deviation from polarity balance
and the flare and/or CME first detection.  Moreover, our systematic 
study shows as many deviations from polarity balance which were 
not related to any 
flare (X or M class) or CME.  So, we conclude that these short-term 
imbalances are not linked to a flare or a CME, but 
rather linked only to the emergence of new fluxes 
(which does not always trigger a 
large coronal activity as expected in a wide range of models).

\begin{acknowledgements}       
       
We thank the referee for careful reading and constructive 
comments on the paper.       
L.M.G.  is grateful to PPARC for postdoctoral funding.  P.D.  and        
C.H.M.  acknowledge financial support from ECOS (France) and SECyT        
(Argentina) through their cooperative science program (A01U04).         
L.v.D.G.  was supported by the Research Fellowship F/02/035 of the        
K.U.  Leuven and by the Hungarian Government grants OTKA T-038013,        
T-032846.   L.M.G., P.D.  and L.v.D.G.  are grateful to the Royal        
Society for a Joint Project award. We acknowledge the SURF for providing data
for use in this publication. The authors thank the MDI, EIT and        
LASCO teams for their data.  SOHO is a project of international        
cooperation between ESA and NASA.         
\end{acknowledgements}

\end{article}       
\end{document}